# AI at work

## Mitigating safety and discriminatory risk with technical standards

Nikolas Becker, Pauline Junginger, Lukas Martinez, Daniel Krupka, Leonie Beining

# AI at work

## Mitigating safety and discriminatory risk with technical standards

Nikolas Becker, Pauline Junginger, Lukas Martinez, Daniel Krupka, Leonie Beining

**TABLE OF CONTENTS**

# Introduction

The use of artificial intelligence (AI) and AI methods in the workplace holds both great opportunities as well as risks to occupational safety and discrimination. In addition to legal regulation, technical standards will play a key role in mitigating such risk by defining technical requirements for development and testing of AI systems. This paper provides an overview and assessment of existing international, European and German standards as well as those currently under development. The paper is part of the research project "ExamAI - Testing and Auditing of AI systems" and focusses on the use of AI in an industrial production environment as well as in the realm of human resource management (HR).

The opportunities of AI in the workplace include e.g., efficiency and quality gains in hiring processes [1], or improvements in safety through AI-based safety functions for robots [2]. On the other hand, we have identified several risks of possible discrimination, particularly in people analytics [3] as well as risks for occupational safety for workers [4].

Therefore, rules ensuring AI systems to be safe and non-discriminatory are required for both the development process as well as the application of AI systems. On the EU level the proposed AI Act and the updated Machinery Regulation (formerly Directive) provide a legal framework for AI regulation and certain aspects of anti-discrimination law, labour law, privacy law, contract and competition law or the civil code do apply. Nevertheless, technical standards will play a key role in shaping the framework conditions and substantiating them. Standards are intended to describe the state of the art and to guide the development and use of technology as well as the design of the associated processes.

This paper is designed to be the starting point for further considerations on the role and integration of standardisation in European AI regulation. We start with further background information on the potentials and challenges of AI systems in the workplace. Then, in section 2, we explain the role of technical standards and their connection to legal regulations. In section 3 the scope of the analysis is specified. Section 4 is the heart of the analysis – a detailed assessment of existing AI standards and those under development. Then, section 5, concludes the paper by highlighting the most promising standards to be revised with AI specific amendments.

# 1.
# AI in workplaces: potentials and challenges

Among numerous other possible applications, artificial intelligence (AI) can be used in factories or warehouses extending the functionality of existing machines such as a robot arm in order to remove people from dangerous physical work like welding. AI thus holds the potential to make classic production automation more flexible [5] and reduce occupational safety and health risks [6].

Collaborating robots (cobots) and Automated Guided Vehicles (AGVs) are two examples of human-machine collaboration in industrial manufacturing that could benefit from the use of AI. Even if cobots and AGVs do not necessarily require the use of AI methods, their functionalities can be supplemented and enhanced by AI. For example, speech recognition can be used to improve human-machine interaction or provide the machine with new capabilities, such as autonomously navigating around obstacles.

AI methods can also be used to develop novel safety features (e.g., warnings of misuse) or optimize existing safety features, as data-driven models can outperform classical software solutions in certain areas such as object recognition [7]. AI-based collision avoidance, for example, could improve the performance of automated driving systems, which could then be deployed more flexibly and widely.

However, the gains in performance and flexibility are offset by risks. When AI methods take on safety-critical functions, it means that errors or failures of AI components could lead to accidents involving property damage or even pose a risk to life and limb. Therefore, if industry intends to benefit from the use of AI methods and manufacturers aim to bring their products to market, all stakeholders must be able to rely on the quality and safety of the system.

The EU Commission's proposals for a horizontal legislative act and for an EU Machinery Regulation address such safety-critical AI applications in cobots and AGVs and classify them as high-risk applications [8]. Like the Machinery Directive, the new regulation requires that machines that use AI to implement safety functions undergo a conformity assessment before being placed on the market within the EU [9]. Due to various characteristics of AI and given the lack of experience that currently exists in dealing with safety-related AI applications, the Commission considers the risk posed by machinery equipped with AI-based safety functions to be particularly high. Accordingly, the Regulatory Proposal/Machinery Regulation requires that the AI component be tested by a notified body and removes the option of internal manufacturing control for the time being. Market surveillance will continue to have the responsibility of monitoring the compliance of products already on the market with safety requirements and the compliance of operators [10].

# 2. The role of technical standards

In contrast to legal regulations, standards are usually not legally binding and are traditionally created by non-state actors. Nonetheless, legislators can create a direct link between standards and law. For example, in the European Union (EU) "harmonized standards" which specify European law (directives and regulations) for products, production processes or services are officially published in the Official Journal of the European Union. According to the "New Approach" established in 1985 and reformed in 2008 with the New Legislative Framework, the full application of harmonized standards is based on conformity with the corresponding essential requirements of the relevant EU directives – the so-called "presumption of conformity" [11,12]. With regard to the EU Machinery Directive (Directive 2006/42/EC), which contains requirements for the safety of machinery, the list of harmonized standards includes, for example, 529 entries [13].

In German law, moreover, various terms in the area of product safety and liability indicate that compliance with standards by manufacturers, system integrators and distributors will positively be taken into account when assessing liability issues [14]. These legal terms include, according to the so called three-stage theory of the German Federal Constitutional Court [15], the "generally accepted rules of technology", the "state of the art", and the more advanced "state of science and technology." The adoption of appropriate standards can therefore mean greater legal certainty for both manufacturers and users of AI systems.

The development of standards is usually a lengthy process that is based on consensus of the general public. In order to make knowledge from research quickly available standardisation organisations also publish specifications and other kinds of documents that only require consensus of the authors involved. By example, the German Institute for Standardization (DIN) publishes specifications under the designation DIN SPEC. DIN SPECs can act as a precursor to a proper DIN standard. On the international level, Technical Specifications (TS) are the equivalent published by the International Organization for Standardization (ISO). Furthermore, ISO publishes Technical Reports (TR). However, Technical Reports have no normative character and are only informative.

[15]

Schneller Brüter, Kalkar I, order of 8 August 1978 - 2 BvL 8/77, Federal Constitutional Court.

# 3.
# Scope of the analysis:
# AI, safety, and testing

The starting point of our consideration are eleven use cases that were identified in the context of the research project "ExamAI – Testing and Auditing of AI systems", funded by the AI Observatory of the German Federal Ministry of Labour and Social Affairs. They are situated in two fields of application: AI systems in human resources and talent management, and AI systems in production automation.

Field of application 1 - AI systems in human resources and talent management:
1. Automated suggestion systems on HR platforms
2. Personality assessment via CV/structured entry or video
3. AI-based background checks
4. HR department chatbot
5. Internal job profile matching
6. Prediction of job termination readiness
7. Automatic work assignment for gig workers

Field of application 2 - AI systems in production automation:
1. Intelligent cobot installs air conditioning compressor incorrectly
2. Intelligent cobot injures worker in the eye
3. Discrimination in route planning of automated guided vehicle (AGV) and forklift drivers
4. Automated guided vehicle (AGV) hits worker

The research project aims at specifying requirements for effective testing and auditing procedures that make the use of AI in the working environment safe and non-discriminatory [16]. They will be designed analogously to conventional testing and auditing procedures in software development [17,18].

We focus on standards that concern the basic safety of software and machines (in particular functional safety), AI-specific standards, and standards for the testing of software systems.

In our overview, we consider standards that are of practical relevance to the German or European market. This includes standards published by relevant German, European, and international standards organizations. In addition, our overview entails standards by the IEEE Standards Association (IEEE SA), which is part of the global professional association of engineers in electrical engineering and information technology IEEE and as such enjoys high international recognition. Thus, we included standards by the following organisations:

- German Institute for Standardization (DIN)
- German Commission for Electrical, Electronic & Information Technologies of DIN and VDE (DKE)
- European Committee for Standardization (CEN)
- European Committee for Electrotechnical Standardization (CENELEC)
- European Telecommunications Standards Institute (ETSI)
- International Organization for Standardization (ISO)
- International Electrotechnical Commission (IEC)
- International Telecommunication Union (ITU)
- Institute of Electrical and Electronics Engineers Standards Association (IEEE SA)

Figure 1 illustrates the level at which the standardization organizations operate and outlines their thematic responsibilities. Due to their interdisciplinary nature various standardization projects at the international level are dealt with in joint technical committees (JTC) of ISO and IEC. In this context, Sub Committee 42 of JTC 1 (JTC 1/SC 42) has been dealing with standardization in the field of AI since 2018 [19]. At European Level, the JTC AI, a joint body of CEN and CENELEC, was formed in 2021 in order to bundle corresponding standardization activities in the field of AI [20]. This includes the adoption of corresponding standards of ISO/IEC JTC 1/SC 42 as a European standard (EN) as well as the development of harmonized standards for the design of a European AI regulation.

[19]
Information on ISO/IEC JTC 1/SC 42 Artificial Intelligence, Revision November 2019. (accessed July 26, 2021)

[20]
Information on New CEN-CLC/JTC on Artificial Intelligence, CEN-CENELEC Focus Group Report: Road Map Report on Artificial Intelligence, version 2020-09.

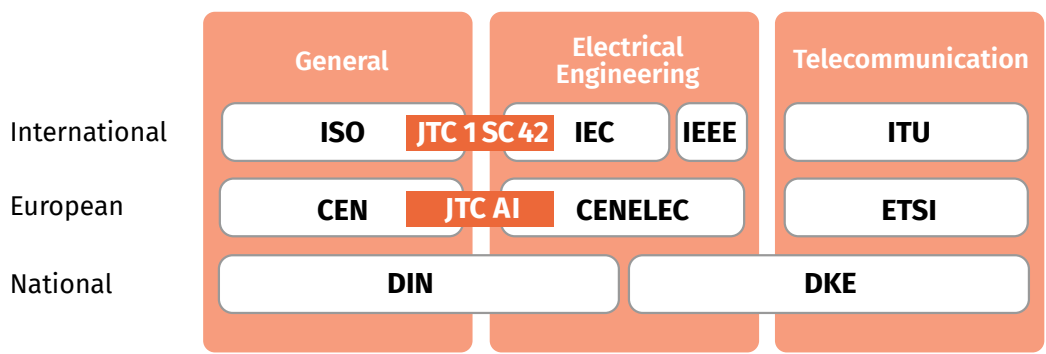

Figure 1: Overview of standardisation organisations

# 4.
# Status of standardisation

Standards from the areas of production automation (Section 3.1) human resources and talent management (Section 3.2) and testing and auditing (Section 3.3) are relevant to the ExamAI project. Several standards are also listed here which do not yet have any AI-specific requirements, but which could be expanded to include such requirements in the future. The affiliation to the application areas is shown in Figure 2. This overview of standards is intended to show only the most relevant standards in relation to the use cases considered and their interrelationships.
Both standards that have already been published and those that are currently still under development were taken into account. These standards are marked accordingly in Figure 2.

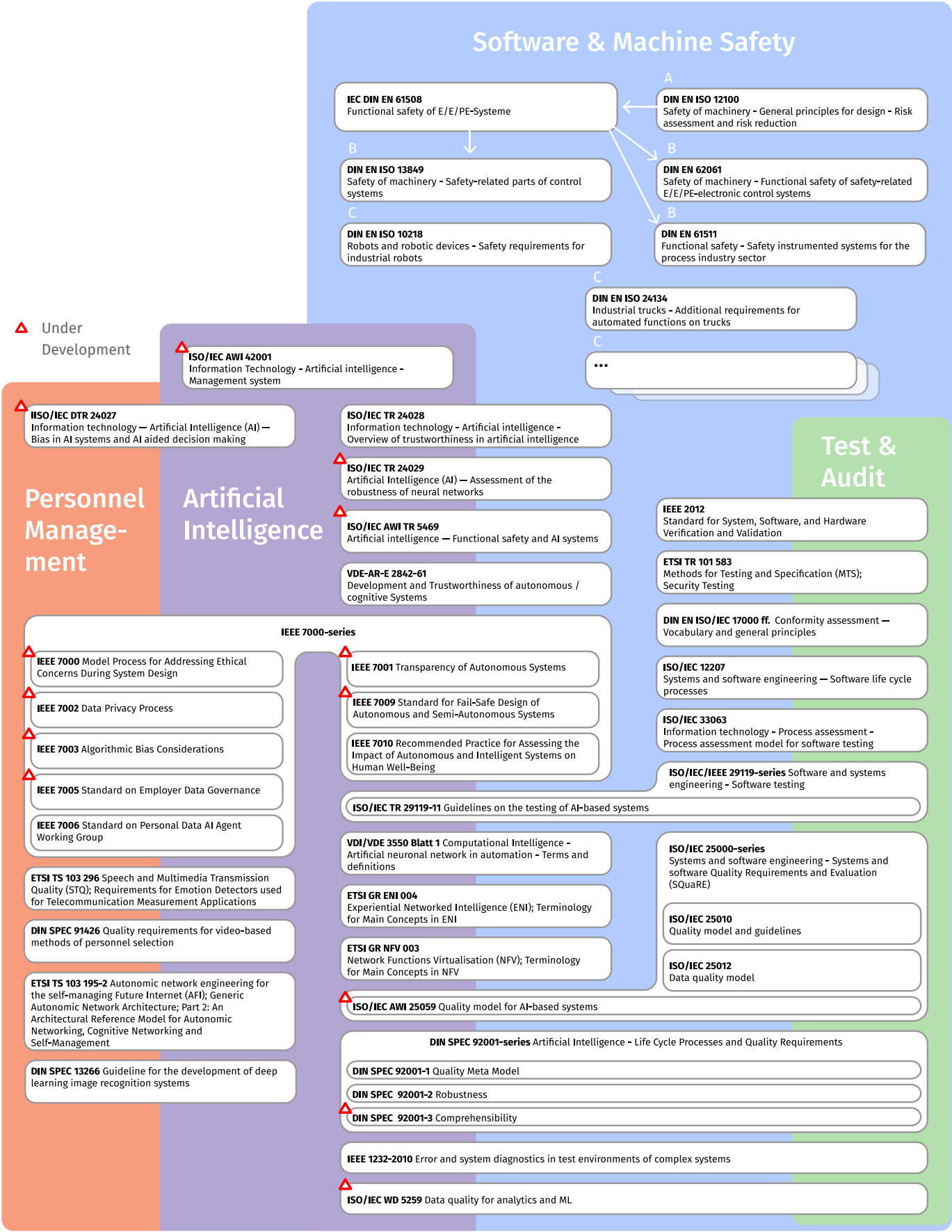

Figure 2: Standards for AI in the workplace

## 4.1. AI in production automation

There are hundreds of standards that set requirements for software and machine safety. With regard to the ExamAI use cases for production automation, those that aim to ensure that control systems for safety-relevant functions function reliably - functional safety - are of particular importance. This includes corresponding risk-reducing measures, but not issues of fire protection, occupational health and safety, or information security (cyber security), which deals with the confidentiality, availability, and integrity of technical systems, i.e., is aimed in particular at protection against external attacks. Specifications for information security are made, for example, by the ISO/IEC 27000 series or IEC 62443, but these will not be considered below.

With regard to machine safety, standards are divided into basic safety standards (type A standards), safety group standards (type B1 and B2 standards) and machine-specific technical standards (type C standards). Type A standards are general in nature, covering basic concepts, design principles and general aspects that affect numerous types of machinery. Type B standards address a safety aspect or type of protective device that affects several or a number of similar machines. Type C standards deal with detailed safety requirements for a group of similar machines, such as industrial trucks in DIN EN ISO 24134, which is why they are of the greatest practical importance [21].

The only type A standard is ISO 12100 *Safety of machinery* and describes general design principles for machines as well as approaches to their risk assessment and reduction. It defines the three basic levels for risk reduction: 1. inherent safety, 2. technical protective measures and supplementary protective devices, and 3. user information. However, it only deals with software with regard to software-controlled safety functions, such as software-controlled light barriers. Otherwise, ISO 12100 refers to IEC 61508 for questions concerning the standardization of software.

IEC 61508 *Functional safety of safety-related electrical/electronic/programmable electronic systems* is thus the basic standard for functional hardware and software safety. It describes the state of the art of safety-related systems containing electrical, electronic or programmable electronic systems. Although IEC 61508 is not yet a harmonized European standard, since it is internationally accepted, it is nevertheless frequently used as a reference when no harmonized standard is applicable. It provides basic requirements for the „entire safety life cycle", which includes design, planning, development, realization, commissioning, maintenance, modification up to decommissioning/ de-installation. Among other things it defines a risk matrix with two dimensions, se-

[21]

G. Steiger, "ISO/TR 22100-1 – Wegweiser für die effektive Nutzung von Typ-A-, Typ-B- und Typ-C-Normen zur Maschinensicherheit," Maschinensicherheit online, Berlin: n.d.(accessed 11 March 2021) verity of damage and probability of its occurrence, according to which Safety Integrity Levels (SIL) 1 to 4 are defined. The use of AI functions for error correction is „neither recommended nor not recommended" at SIL 1 (the lowest level) (Table A.2). Nevertheless, IEC 61508- 7:2010 Section C.3.9 already mentions artificial intelligence as a very efficient way to predict, correct, maintain and control faults.

Various Type B standards are derived from IEC 61508, of which the Type B standards IEC 62061 *Safety of machinery - Functional safety of safety-related electrical, electronic and programmable electronic control systems and* EN ISO 13849 *Safety of machinery - Safety-related parts of control systems* are of particular importance for machine safety. Also, worth mentioning is IEC 61511, which decribes functional safety in automation and control systems - but only for the process industry. It could be of importance for issues relating to the networked factory. More relevant is EN ISO 13849, which also provides safety requirements for software as well as a guideline for the design and integration of safety-related parts of machine controls.

So far, IEC 61508 does not consider the specifics of AI systems, let alone learning systems. The standard ISO/IEC TR 24028 *Overview of trustworthiness in artificial intelligence* from 2020 does address AI machines, but in the case of safety-critical functions it only refers to implementing classic safety functions such as light barriers as defined by IEC 61508.

Another approach to extending IEC 61508 to include AI aspects is being developed in the WG3 working group of the ISO/IEC JTC 1/SC 42 with ISO/IEC AWI TR 5469 *Artificial intelligence - Functional safety and AI systems*. This standard is intended not only to address the control of AI systems, but also to describe the development of AI-based safety-related functions. The associated idea of probabilistic risk acceptance criteria is as promising as it is controversial.

Furthermore, at the German level, the application rule VDE-AR-E 2842-61 *Development and Trustworthiness of autonomous / cognitive Systems* represents an attempt to extend the view of a machine as mechanically interconnected parts, as specified in ISO 12100, and to take the entire socio-technical system into consideration. In addition to these last three documents, all of which do not have the status of an adopted standard and are still under development, there are so far primarily standards which define terminologies and concepts in connection with AI (VDI/VDE 3550, ETSI GR ENI 004, ETSI GR NFV 003).

One series of standards that addresses challenges related to bias and other ethical aspects in the development of autonomous systems is the IEEE P7000 series *Ethics in Action in Autonomous and Intelligent Systems*. IEEE Std 7010 *IEEE Recommended Practice for Assessing the Impact of Autonomous and Intelligent Systems on Human Well-Being* has been published from this series as the only one of 13 planned standards to date and measures the impact of autonomous and intelligent systems on humans or human well-being. Other planned standards in the series with relevance for the application field of production automation are, for example, P7001 *Transparency of Autonomous Systems* or P7009 *Standard for Fail-Safe Design of Autonomous and Semi-Autonomous Systems.*

## 4.2. AI in HR and talent management

In the context of human resources and talent management, there are only a few standardized or normed requirements for AI systems. Requirements for freedom from undue discrimination (bias) have so far been addressed, among other things, by the IEEE 7000 series *Model Process for Addressing Ethical Concerns During System Design*, which is currently being developed and has already been mentioned in the application field of production automation. It is dedicated to the development of ethical standards for the development of autonomous and intelligent systems. The standards P7000 *Model Process for Addressing Ethical Concerns During System Design*, P7002 *Data Privacy Process*, P7003 *Algorithmic Bias Considerations*, P7005 Standard on Employer Data Governance and P7006 *Standard on Personal Data AI Agent Working Group*, which have not yet been published, are particularly relevant for the application field of human resources and talent management.

Another relevant standard for the application field of personnel and talent management is DIN SPEC 91426 *Quality requirements for video-based methods of personnel selection*. As requirements for the product features and functionalities of these methods, the SPEC also names, among other things, the input and output files to be used as critical components. These must be selected without prejudice or stereotype and checked by trained personnel using suitable data analysis procedures to identify errors (bias).

In addition to this standard, which is currently available as a SPEC, the ISO/IEC DTR 24027 *Bias in AI systems and AI aided decision* making should be mentioned as a relevant project. This standard, which is still under development, is explicitly dedicated to dealing with bias-related risks in the development of AI and ADM systems.

In addition, there are a number of standards such as DIN SPEC 13266, which sets requirements for the development of image recognition systems, and ETSI TS 103 296, which is dedicated to emotion recognition.

## 4.3. Testing and auditing of AI systems

With regard to (software) testing, there are two established and comprehensive series of standards, ISO/IEC/IEEE 29119 and ISO/IEC 25000. The ISO/IEC 29119 series claims to be valid for software testing of any form in any organization. It includes test process descriptions that define the software test processes at the organizational level, the test management level and the dynamic test levels. In addition to dynamic testing, it supports functional and non-functional testing, manual and automated testing, and scripted and unscripted testing. Part 11 of this series (ISO/IEC TR 29119-11 *Testing of AI-based systems*), published in 2020, also addresses the testing of AI systems.

The comprehensive framework for *System and Software Quality Requirements and Evaluation (SQuaRE)* in the ISO/IEC 25000 series also defines requirements for functional and non-functional tests. The ISO/IEC 2501n standards (Quality Model Division) describe detailed quality models, for example in ISO/IEC 25010 for system and software quality and in ISO/IEC 25012 for data quality. The standards in ISO/IEC 2502n contain a reference model for measuring software quality and instructions for its application. ISO/IEC AWI 25059 *Quality model for AI-based systems*, which is currently being developed, is intended to formulate specific requirements for the quality of AI-based systems.

With the *AI quality metamodel* defined in DIN SPEC 92001, there is also an attempt to provide a framework explicitly for testing AI systems. Based on this, DIN SPEC 92001-2 sets requirements for robustness. DIN SPEC 92001-3 is a standard in progress that formulates requirements for the comprehensibility of AI systems. As DIN SPECs, these rules will not have the status of a standard, but can serve as a basis for later standardization.

At the international level, ISO/IEC 42001 *Information Technology - Artificial intelligence - Management System* is a standard that aims to provide a recommended course of action for the design, implementation and maintenance of AI-based management systems in organizations. The standard is intended to help organizations to use AI responsibly and to take appropriate account of ethical aspects that are relevant in the context of AI. This should also increase the trust of consumers in AI-based systems.

ExamAI – KI Testing & Auditing

In addition, the new series ISO / IEC WD 5259 *Data quality for analytics and ML* addresses the topic of data quality in the context of training and evaluation data. Four standards are currently planned in this series. Among other things, they will describe criteria for the quality of data and a standardized procedure for data processing.

The ISO / IEC TS 4213 *Information Technology - Artificial Intelligence - Assessment of machine learning classification performance standard,* which is currently in preparation, will also specify methods for measuring the classification performance of ML-based models, systems and algorithms.

Finally, the proposal ISO / IEC TR 24029-2 *Information Technology - Artificial Intelligence (AI) - Assessment of the robustness of neural networks - Part 2: Methodology for the use of formal methods* should be mentioned. It is intended to describe formal methods for the assessment of the robustness of neural networks.

It became clear that a whole range of standards and norms are under development in connection with the testing and auditing of AI systems, but many of them are still in early stages of development (cf. "Stage" in the tabular overview).

# 5. Conclusion

Current standards do not provide manufacturers and system integrators with clear instructions on how AI can be integrated in the application areas of human resources and production automation in a safe, non-discriminatory or generally legally compliant manner. The testing of corresponding AI systems is also insufficiently described. Nevertheless, there are already technical reports, such as ISO/IEC TR 29119-11 in particular, which could be further developed into standards, as well as several standardization projects (ISO/IEC AWI 25059, ISO/IEC AWI 42001, etc.) which have begun the corresponding work.

In view of the large number of relevant Type C standards - 421 for the Machinery Directive alone - it will not be possible to implement an extension of these to include AI aspects in the short term. A revision of the 104 harmonized Type B standards appears more realistic. Should there be a fundamental paradigm shift in the assessment of risk acceptance for the use of AI-based safety functions, however, this would first have to be established at type A level in IEC 61508. In particular, this raises the question of adapting the safety integrity levels [22].

ISO 12100, on the other hand, will probably not need to be revised. According to the recently published document ISO/TR 22100-5:2021 *Safety of machinery - Relationship with ISO 12100 - Part 5: Implications of artificial intelligence machine learning*, the methodology of ISO 12100 is also suitable for addressing the risks of embedded AI systems, as long as the AI only performs individual tasks.

In the standardized test procedures, there is a particular need for action with regard to greater consideration of fairness, non-discrimination and data protection as non-functional or extra-functional requirements. This could be done within the framework of the ISO/IEC 25000, ISO/IEC 29119 and/or DIN SPEC 92001 series of standards. The revision could be based on the IEEE P7000 series.

[22]
J. Braband, H. Schäbe, "On Safety Assessment of Artificial Intelligence," arXiv preprint arXiv:2003.00260, 2020.

# Annex 1:
# Table of standards

| Standard | Stage [23] | Title |
| --- | --- | --- |
| IEC DIN EN 61508 | Published | Functional safety of electrical/electronic/programmable electronic safety-related systems |
| DIN EN ISO 12100 | Published | Safety of machinery - General principles for design - Risk assessment and risk reduction |
| DIN EN ISO 13849 | Published | Safety of machinery - Safety-related parts of control systems |
| DIN EN 62061 | Published | Safety of machinery - Functional safety of safety-related electrical, electronic and programmable electronic control systems |
| DIN EN ISO 10218 | Published | Robots and robotic devices - Safety requirements for industrial robots |
| DIN EN 61511 | Published | Functional safety - Safety instrumented systems for the process industry sector |
| DIN EN ISO 24134 | Published | Industrial trucks - Additional requirements for automated functions on trucks |
| ISO/IEC AWI 42001 | 20.00 Preparatory | Information Technology - Artificial intelligence - Management system |

[23]
International harmonized stage codes https://www.iso.org/stage-codes.html

| ISO/IEC DTR 24027 | 30.60 Committee | Information technology — Artificial Intelligence (AI) — Bias in AI systems and AI aided decision making |
| --- | --- | --- |
| ISO/IEC TR 24028 | Published | Information technology - Artificial intelligence - Overview of trustworthiness in artificial intelligence |
| ISO/IEC TR 24029 | Teil 1: 60.60 Published Teil 2: 10.99 Proposal | Artificial Intelligence (AI) — Assessment of the robustness of neural networks |
| ISO/IEC AWI TR 5469 | 10.99 Proposal | Artificial intelligence — Functional safety and AI systems |
| ISO/IEC TS 4213 | 20.00 Preparatory | Information technology — Artificial Intelligence — Assessment of machine learning classification performance |
| IEEE 2012 | Published | Standard for System, Software, and Hardware Verification and Validation |
| VDE -AR-E 2842-61 | Published | Development and trustworthiness of autonomous/cognitive systems |
| IEEE 7000 | Project/Draft | Model Process for Addressing Ethical Concerns During System Design |
| IEEE 7001 | Project/Draft | Transparency of Autonomous Systems |
| IEEE 7002 | Project/Draft | Data Privacy Process |
| IEEE 7003 | Project/Draft | Algorithmic Bias Considerations |
| IEEE 7005 | Project/ Draft | Standard on Employer Data Governance |
| IEEE 7009 | Project/ Draft | Standard for Fail-Safe Design of Autonomous and Semi-Autonomous Systems |
| IEEE 7010 | Published | Recommended Practice for Assessing the Impact of Autonomous and Intelligent Systems on Human Well-Being |
| DIN EN ISO/IEC 17000 ff. | Published | Conformity assessment - Vocabulary and general principles |
| ISO/IEC/IEEE 12207 | Published | Systems and software engineering - Software life cycle processes |
| ISO/IEC 33063 | Published | Information technology - Process assessment - Process assessment model for software testing |
| ISO/IEC/IEEE 29119-series | Published | Software and systems engineering - Software testing |
| ISO/IEC TR 29119-11 | Published | Part 11: Guidelines on the testing of AI-based systems |
| ETSI TS 103 296 | Published | Speech and Multimedia Transmission Quality (STQ); Requirements for Emotion Detectors used for Telecommunication Measurement Applications |
| VDI/VDE 3550 Blatt 1 | Published | Computational Intelligence - Artificial neuronal network in automation - Terms and definitions |
| ISO/IEC 25000-series | Published | Systems and software engineering - Systems and software Quality Requirements and Evaluation (SQuaRE) |

ExamAI – KI Testing & Auditing| ISO/IEC 25010 | Published | System and software quality models |
|---|---|---|
| ISO/IEC 25012 | Published | Data quality model |
| ISO/IEC AWI 25059 | 20.00 Preparatory | Quality model for AI-based systems |
| DIN SPEC 91426 | Published | Quality requirements for video-based methods of personnel selection |
| ETSI GR ENI 004 | Published | Experiential Networked Intelligence (ENI); Terminology for Main Concepts in ENI |
| ETSI GR NFV 003 | Published | Network Functions Virtualisation (NFV); Terminology for Main Concepts in NFV |
| ETSI TS 103 195-2 | Published | Autonomic network engineering for the self-managing Future Internet (AFI); Generic Autonomic Network Architecture; Part 2: An Architectural Reference Model for Autonomic Networking, Cognitive Networking and Self-Management |
| DIN SPEC 13266 | Published | Guideline for the development of deep learning image recognition systems |
| DIN SPEC 92001-series | Published | Artificial Intelligence - Life Cycle Processes and Quality Requirements<br><br>92001-1 Quality Meta Model<br>92001-2 Robustness<br>92001-3 Comprehensibility [Preliminary] |
| IEEE 1232 | Published | Artificial Intelligence Exchange and Service Tie to All Test Environments (AI-ESTATE) |
| ISO/IEC WD 5259 | 20.00 Preparatory | Data quality for analytics and ML |

# Imprint

### About "ExamAI – Testing and Auditing of AI systems"
In the consortium project "ExamAI – Testing and Auditing of AI systems", led by the German Informatics Society, an interdisciplinary team of (socio-)computer scientists, software engineers as well as law and political scientists examines meaningful control and testing procedures for AI systems. The project focuses on two concrete application areas: human-machine cooperation in industrial production and AI systems in personnel and talent management. You can find out more about the results of the project at https://testing-ai.gi.de
The project is funded by the Federal Ministry of Labour and Social Affairs' Policy Lab Digital, Work & Society as part of the Observatory for Artificial Intelligence in Work and Society (AI Observatory) project. The Policy Lab Digital, Work & Society is a new organisational unit within the "Digitalisation and the Labour Market Department" at the Federal Ministry of Labour and Social Affairs. It is a new, interdisciplinary and agile organisational unit that observes technological, economic and social trends and helps to shape change together with academia, business and social partners. The views expressed in this paper do not necessarily represent the official positions of the ministry.